\documentstyle[aps,epsfig]{revtex}
\newcommand{\lsim}{\mathrel{\rlap{\lower4pt\hbox{\hskip0pt$\sim$}}
\raise1pt\hbox{$<$}}}
\newcommand{\gsim}{\mathrel{\rlap{\lower4pt\hbox{\hskip0pt$\sim$}}
\raise1pt\hbox{$>$}}}
\newcommand{\sfrac}[2]{\mbox{\footnotesize $\frac{#1}{#2}$}}
\begin{document}
\twocolumn

\title{$T$-dependence of pseudoscalar and scalar correlations}
%
\author{P. Maris,\footnotemark[2] C.D. Roberts,\footnotemark[1]
S.M. Schmidt\footnotemark[1] and P.C.~Tandy\footnotemark[2]\vspace*{0.4em}}
\address{
\footnotemark[2]Center for Nuclear Research, Department of Physics, 
Kent State University, Kent OH 44242-0001\vspace*{0.2em}\\
\footnotemark[1]Physics Division, 
Argonne National Laboratory, Argonne IL
60439-4843\\[0.6\baselineskip] 
\parbox{140mm}{\rm \hspace*{1.0em} 
Inhomogeneous pseudoscalar and scalar Bethe-Salpeter equations solved using a
renormalisation-group-improved rainbow-ladder truncation exhibit bound state
poles below and above $T_c$, the critical temperature for chiral symmetry
restoration.  Above $T_c$ the bound state amplitudes are identical, as are
the positions and residues of the pseudoscalar and scalar poles in the
vertices.  In the chiral limit the $\pi^0\to \gamma\gamma$ coupling vanishes
at $T_c$, as do $f_\pi$, $m_\sigma$, $g_{\sigma\pi\pi}$.  For light
current-quark masses the $2 \pi$ decay channel of the isoscalar-scalar meson
remains open until very near $T_c$, and the widths of the dominant pion decay
modes remain significant in the vicinity of the
crossover.\\[0.4\baselineskip]
Pacs Numbers: 11.10.Wx, 11.30.Rd, 12.38.Mh, 24.85.+p
}}
%
%
\maketitle



\section{Introduction}
Numerical simulations of $2$-light-flavour lattice-QCD, and the study of
models that accurately describe dynamical chiral symmetry breaking and the
$\pi$-$\rho$ mass-difference at $T=0$ both indicate that a quark-gluon plasma
(QGP) is reached via a second order transition at a critical temperature $T_c
\approx 0.15\,$GeV\cite{echaya2,hirschegg97,trento99}.  This QGP existed as a
stage in the early evolution of the universe and its terrestrial recreation
is a primary goal of current-generation ultrarelativistic heavy-ion
experiments.  The plasma is characterised by the free propagation of quarks
and gluons over distances $\sim 10$-times larger than the proton.  However,
its formation can only be observed indirectly by searching for a modification
of particle yields and/or hadron properties in the debris of the collisions.

The plasma phase is also characterised by chiral symmetry restoration.
Hence, since the properties of the pion (mass, decay constant, other vertex
residues, etc.) are tied to the dynamical breaking of chiral symmetry, an
elucidation of the $T$-dependence of these properties is important;
particularly since a prodigious number of pions is produced in heavy ion
collisions.  Also important is understanding the $T$-dependence of the
properties of the scalar analogues and chiral partners of the pion in the
strong interaction spectrum.  For example, should the mass of a putative
light isoscalar-scalar
meson\cite{mikescalars,mikescalar2,mikenew,jacquesscalar} fall below
$2\,m_\pi$, the strong decay into a two pion final state can no longer
provide its dominant decay mode.  In this case electroweak processes will be
the only open decay channels below $T_c$ and the state will appear as a
narrow resonance.  Analogous statements are true of isovector-scalar mesons.

The Dyson-Schwinger equations (DSEs)\cite{cdragw} provide a nonperturbative,
continuum framework for analysing quantum field theories and the class of
rainbow-ladder truncation DSE models yields a qualitative understanding of
the thermodynamic properties of the QGP phase transition at $T\neq
0$\cite{echaya} (and also at $\mu \neq 0$\cite{echaya,Blaschke:1999,dqppr}).
For example, using a simple element of this class the pressure's slow
approach to its ultrarelativistic limit, which is observed in numerical
simulations of lattice-QCD\cite{Engels:1997}, could be attributed to a
persistence into the QGP of nonperturbative effects in the quark's vector
self energy\cite{basti97}.\footnote{$T=0$ DSE studies characteristically
yield momentum-dependent vector and scalar quark self energies that remain
large until $k^2=1\,$GeV$^2$; e.g., Fig.~8 of Ref.~\protect\cite{mr97} and
Ref.~\cite{kisslinger}.  These features are also observed in lattice
simulations\protect\cite{Skullerud:1999}.}
Also, using a more sophisticated model, the present incompatibility between
lattice estimates of the critical exponents\cite{echaya2} was identified as
likely an artefact of working too far from the chiral limit\cite{Holl:1999}.

The calculation of the $T$-dependence of hadron properties has hitherto used
only simple members of this class\cite{Blaschke:1999,Maris:1998,peterNew}.
Herein we employ a version that provides for renormalisability and the
correct one-loop renormalisation group evolution of scale-dependent matrix
elements, and focus on scalar and pseudoscalar properties.  Mesons appear as
simple poles in $3$-point vertices.  Importantly, however, these vertices
also provide information about the persistence of correlations away from the
bound state pole; e.g., Refs.~\cite{Maris:1999,jacques}, which can be useful
in studying the $T$-evolution of a system with deconfinement.  Consequently
our starting point is the inhomogeneous Bethe-Salpeter equation (BSE).  We
then employ the homogeneous equation when appropriate and useful.

In Sec.~\ref{sec2} we describe the $T\neq 0$ quark DSE and inhomogeneous
BSEs, and introduce our model.  This also serves to make clear our notation.
Section~\ref{sec3} reports our results for the $T$-dependence of scalar and
pseudoscalar correlations while Sec.~\ref{sec4} is a brief recapitulation and
epilogue.  An appendix contains selected formulae.

\section{Dyson-Schwinger and Bethe-Salpeter equations}
\label{sec2}
The renormalised quark DSE is 
\begin{eqnarray}
S^{-1}(p_{\omega_k})  & := & i\vec{\gamma}\cdot \vec{p} \,A(p_{\omega_k})
+ i\gamma_4\,\omega_k \,C(p_{\omega_k} ) + B(p_{\omega_k} ) 
\end{eqnarray}
\begin{eqnarray}
& = & Z_2^A
\,i\vec{\gamma}\cdot \vec{p} + Z_2 \, (i\gamma_4\,\omega_k + m_{\rm bm})\, +
\Sigma^\prime(p_{\omega_k} )\,,
\label{qDSE} 
\end{eqnarray}
where $p_{\omega_k}:= (\vec{p},\omega_k)$ with $\omega_k= (2 k + 1)\,\pi T$
the fermion Matsubara frequency, and $m_{\rm bm}$ is the Lagrangian
current-quark bare mass.  The regularised self energy is
\begin{equation}
\label{sigmap}
\Sigma^\prime(p_{\omega_k}) = i\vec{\gamma}\cdot
  \vec{p}\,\Sigma_A^\prime(p_{\omega_k} ) +
  i\gamma_4\,\omega_k\,\Sigma_C^\prime(p_{\omega_k} ) +
  \Sigma_B^\prime(p_{\omega_k})\,, 
\end{equation}
\vspace*{-2.5\baselineskip}

\begin{eqnarray}
\nonumber \Sigma_{\cal F}^\prime(p_{\omega_k}) & =&
\int_{l,q}^{\bar\Lambda}\,
\case{4}{3}\,g^2\,D_{\mu\nu}(\vec{p}-\vec{q},\omega_k-\omega_l) \\ 
&& \times \,\case{1}{4}{\rm tr}\left[{\cal P}_{\cal F} \gamma_\mu
S(q_{\omega_l})\Gamma_\nu(q_{\omega_l};p_{\omega_k})\right]\,,
\label{regself}
\end{eqnarray}
where: ${\cal F}=A,B,C$; $A,B,C$ are functions of $(|\vec{p}|^2,\omega_k^2)$;
\begin{equation}
{\cal P}_A:= -(Z_1^A/|\vec{p}|^2)i\vec{\gamma}\cdot \vec{p}\,,
{\cal P}_B:= Z_1 \,, 
{\cal P}_C:= -(Z_1/\omega_k)i\gamma_4\,;
\end{equation}
and $\int_{l,q}^{\bar\Lambda}:=\, T
\,\sum_{l=-\infty}^\infty\,\int^{\bar\Lambda}d^3q/(2\pi)^3$, with
$\int^{\bar\Lambda}$ representing a translationally invariant regularisation
of the integral and $\bar\Lambda$ the regularisation mass-scale.  The
renormalised self energies are
\begin{equation}
\label{renself}
\begin{array}{rcl}
{\cal F}(p_{\omega_k};\zeta) & = & 
\xi_{\cal F} + \Sigma_{\cal F}^\prime(p_{\omega_k};{\bar\Lambda})
    - \Sigma_{\cal F}^\prime(\zeta^-_{\omega_0};{\bar\Lambda})\,,
\end{array}
\end{equation}
$\zeta$ is the renormalisation point, $(\zeta^-_{\omega_0})^2 := \zeta^2 -
\omega_0^2$, $\xi_A = 1 = \xi_C$, and $\xi_B=m_R(\zeta)$.

\subsection{The Model}
$\Gamma_\nu(q_{\omega_l};p_{\omega_k})$ in Eq.~(\ref{regself}) is the
renormalised dressed-quark-gluon vertex.  It is a connected, irreducible
$3$-point function that should not exhibit light-cone singularities in
covariant gauges~\cite{hawes}.  A number of {\it Ans\"atze} with this
property have been proposed and it has become clear that the judicious use of
the rainbow truncation
\begin{equation}
\Gamma_\nu(q_{\omega_l};p_{\omega_k}) = \gamma_\nu
\end{equation}
in Landau gauge provides phenomenologically reliable results so we employ it
herein.  A mutually consistent constraint is
\begin{equation}
Z_1 = Z_2\;\;{\rm and}\;\;Z_1^A = Z_2^A\,.
\end{equation}
The rainbow truncation is the leading term in a $1/N_c$ expansion of
$\Gamma_\nu(q_{\omega_l};p_{\omega_k})$.

$D_{\mu\nu}(p_{\Omega_k})$ is the renormalised dressed-gluon propagator
\begin{equation}
g^2 D_{\mu\nu}(p_{\Omega_k}) = 
P_{\mu\nu}^L(p_{\Omega_k} ) \Delta_F(p_{\Omega_k} ) + 
P_{\mu\nu}^T(p_{\Omega_k}) \Delta_G(p_{\Omega_k}  ) \,,
\end{equation}
\begin{eqnarray}
P_{\mu\nu}^T(p_{\Omega_k}) & := &\left\{
\begin{array}{ll}
0, &  \mu\;{\rm and/or} \;\nu = 4,\\
\displaystyle
\delta_{ij} - \frac{p_i p_j}{p^2}, &  \mu,\nu=i,j\,=1,2,3\;,
\end{array}\right.\\
P_{\mu\nu}^L(p_{\Omega_k}) & + & P_{\mu\nu}^T(p_{\Omega_k}) = 
\delta_{\mu\nu}- \frac{p_\mu p_\nu}{p^2}\,,
\end{eqnarray}
$\mu,\nu= 1,\ldots, 4$, and $\Omega_k:= 2\pi k T$.  A Debye mass for the
gluon appears as a $T$-dependent contribution to $\Delta_F$.

The ultraviolet behaviour of the kernel in Eq.~(\ref{regself}) is fixed by
perturbative QCD because the DSEs yield perturbation theory in the weak
coupling limit.  Our model is defined by specifying a form for the kernel's
infrared behaviour:
\begin{eqnarray}
\label{uvpropf}
\Delta_F(p_{\Omega_k}) & = & {\cal D}(p_{\Omega_k};m_g)\,,\;
\Delta_G(p_{\Omega_k})  =  {\cal D}(p_{\Omega_k};0)\,,\\
\label{delta}
 {\cal D}(p_{\Omega_k};m_g) & := & 
        2\pi^2 D\,\case{2\pi}{T}\delta_{0\,k} \,\delta^3(\vec{p}) 
        + {\cal D}_{\rm M}(p_{\Omega_k};m_g)\,,
\end{eqnarray}
where $D= (0.881\,{\rm GeV})^2$ is a mass-scale parameter and 
\begin{eqnarray}
\nonumber 
\lefteqn{ {\cal D}_{\rm M}(p_{\Omega_k};m_g) = 
\frac{4\pi^2}{\omega^6} D \,s_{\Omega_k} \,
        {\rm e}^{-s_{\Omega_k} /\omega^2}}\\
\label{modelmr}
&& + \frac{8\pi^2\gamma_m}
{\ln
\left[\tau + \left(1+s_{\Omega_k}/\Lambda_{\rm QCD}^2\right)^2\right]}\,
\frac{1-{\rm e}^{- s_{\Omega_k} /(4m_t^2)}}
        {s_{\Omega_k} }\,,
\end{eqnarray}
with $s_{\Omega_k} = p_{\Omega_k}^2 +m_g^2$, $\tau = e^2-1$, $\omega =
1.2\,m_t$, $m_t= 0.5\,$GeV, $\gamma_m= 12/25$, $m_g^2 = (16/5)\, \pi^2 T^2$,
and $\Lambda_{\rm QCD}^{N_f=4}=0.234\,$GeV.  This model, which is motivated
by Refs.~\cite{herman,herman2}, incorporates the one-loop logarithmic
suppression identified in perturbative calculations.  Its parameters were
fixed at $T=0$ by fitting a range of $\pi$ and $K$ meson properties: a
renormalisation point invariant light current-quark mass $\hat
m_{u,d}=5.7\,$MeV, corresponding to $m_R(1\,{\rm GeV})=4.8\,$MeV, gives
$m_\pi=0.14\,$GeV, $f_\pi=0.092\,$MeV.  The model exhibits a second order
chiral symmetry restoring transition at\cite{Holl:1999}
\begin{equation}
T_c=0.15\,{\rm GeV.}
\end{equation}

\subsection{Pseudoscalar Channel}
Ward-Takahashi identities relate the $3$-point vector and axial-vector
vertices to the dressed-quark propagator.  Hence, once a truncation of the
kernel in the quark DSE has been selected, requiring the preservation of
these identities constrains the kernel in the BSE.  This is explored in
Ref.~\cite{axel}, where a systematic procedure for constructing the kernels
is introduced that ensures the order-by-order preservation of these
identities.

Using this procedure the inhomogeneous BSE for the zeroth Matsubara mode of
the isovector $0^{-+}$ vertex consistent with the rainbow truncation of
Eq.~(\ref{qDSE}) is
\begin{eqnarray}
\nonumber
\lefteqn{
\Gamma_{\rm ps}^i(p_{\omega_k};P_0;\zeta)  =  Z_4
\,\case{1}{2}\tau^i\gamma_5} \\
\nonumber 
\lefteqn{- \int_{l,q}^{\bar\Lambda} \,\case{4}{3}\,
g^2 D_{\mu\nu}(p_{\omega_k}-q_{\omega_l})} \\ 
\label{psvtx}
&  \times & \gamma_\mu S(q_{\omega_l}^+)\,\Gamma_{\rm
ps}^i(q_{\omega_l};P_0;\zeta)\, S(q_{\omega_l}^-)\,\gamma_\nu\,,
\end{eqnarray}
where $\{\tau^i$, $i=1,2,3$\} are the Pauli matrices, $q_{\omega_l}^\pm=
q_{\omega_l}\pm P_0/2$, $P_0=(\vec{P},0)$, and $Z_4=Z_4(\zeta,\bar\Lambda)$
is the mass renormalisation constant:
\begin{equation}
m_R(\zeta)\,\Gamma_{\rm ps}^i(p_{\omega_k};P_{\Omega_n};\zeta)
\end{equation}
is renormalisation point independent.

Equation~(\ref{psvtx}) is a dressed-ladder BSE.  At $T=0$ its solution
exhibits poles, and their positions and residues provide a good description
of light vector and flavour-nonsinglet pseudoscalar mesons when $S$ is
obtained from the rainbow quark DSE\cite{mr97,mt99}.  This truncation is
reliable because of cancellations between vertex corrections and crossed-box
contributions at each higher order in the quark-antiquark scattering kernel.

The solution of Eq.~(\ref{psvtx}) has the form (hereafter the
$\zeta$-dependence is often implicit)
\begin{eqnarray}
\nonumber
\lefteqn{\Gamma_{\rm ps}^i(p_{\omega_k};\vec{P}) = }\\
&& \nonumber
\case{1}{2}\tau^i\gamma_5\,\left[i E_{\rm ps} (p_{\omega_k};\vec{P})
+ \vec{\gamma}\cdot\vec{P} \,F_{\rm ps} (p_{\omega_k};\vec{P})
\right. \\
& & \left.  
+\vec{\gamma}\cdot\vec{p}\,\vec{p}\cdot\vec{P}\,
        G^{\|}_{\rm ps} (p_{\omega_k};\vec{P})
+ \gamma_4\omega_k\,\,\vec{p}\cdot\vec{P}\,
G^\perp_{\rm ps} (p_{\omega_k};\vec{P})
\right]\,,
\label{psvtxform}
\end{eqnarray}
where we have neglected terms involving $\sigma_{\mu\nu}$-like contributions,
which play a negligible role at $T=0$\cite{mr97}.
The scalar functions in Eq.~(\ref{psvtxform}) exhibit a simple pole at
$\vec{P}^2+m_\pi^2 = 0$ so that
\begin{equation}
\label{IHVpspole}
\Gamma_{\rm ps}^i(p_{\omega_k};\vec{P}) = 
\frac{r_{\pi}(\zeta)}{\vec{P}^2 + m_\pi^2}\,
\Gamma_\pi^i(p_{\omega_k};\vec{P}) + {\rm regular,}
\end{equation}
where ``regular'' means terms regular at {\it this} pole and
$\Gamma_\pi^i(p_{\omega_k};\vec{P})$ is the canonically normalised, bound
state pion Bethe-Salpeter amplitude:
\begin{eqnarray}
\nonumber
\lefteqn{2 \delta^{ij} \vec{P}  =  
{\rm tr}\,\int_{l,q}^{\bar\Lambda}\,\left\{
\Gamma_\pi^i(q_{\omega_l};-\vec{P}) \, 
\frac{\partial S(q_{\omega_l}^+)}{\partial \vec{P}} 
\Gamma_\pi^j(q_{\omega_l};\vec{P}) \,  S(q_{\omega_l}^-)\right.}\\
&& +
\left. \left.
\Gamma_\pi^i(q_{\omega_l};-\vec{P}) \, S(q_{\omega_l}^+)
\Gamma_\pi^j(q_{\omega_l};\vec{P}) \,
\frac{\partial S(q_{\omega_l}^-)}{\partial \vec{P}} 
\right\}\right|_{\vec{P}^2= -m_\pi^2}\!\!\!,
\end{eqnarray}
with the trace over colour, Dirac and isospin indices, and the residue is
\begin{equation}
\label{rpi}
\delta^{ij}\, ir_{\pi} = Z_4\,{\rm tr}\int_{l,q}^{\bar\Lambda}\,
\case{1}{2}\tau^i\,\gamma_5 \chi_\pi^j(q_{\omega_l};\vec{P})\,,
\end{equation}
where $\chi_\pi(q_{\omega_l};\vec{P}):= S(q_{\omega_l}^+)
\Gamma_\pi(q_{\omega_l};\vec{P})\, S(q_{\omega_l}^-)$ is the unamputated
Bethe-Salpeter wave function.  Substituting Eq.~(\ref{IHVpspole}) into
Eq.~(\ref{psvtx}) and equating pole residues yields the homogeneous pion BSE,
which provides the simplest way to obtain the bound state amplitude.

At $T=0$, $r_{\pi}(\zeta)$ is the gauge-invariant pseudoscalar projection of
the pion Bethe-Salpeter wave function at the origin in configuration space;
i.e, it is a field theoretical analogue of the ``wave function at the
origin,'' which describes the decay of bound states in quantum mechanics.

The residue of the pion pole in the axial-vector vertex is the pion decay
constant:
\begin{equation}
\label{fpi}
\delta^{ij}\, \vec{P}\,f_\pi  = 
Z_2^A\,{\rm tr}\int_{l,q}^{\bar\Lambda}\,
\case{1}{2}\tau^i\,\gamma_5 \vec{\gamma}\,
\chi_\pi^j(q_{\omega_l};\vec{P})\,.
\end{equation}
It is the gauge-invariant pseudovector projection of Bethe-Salpeter wave
function at the origin, which completely determines the strong interaction
contribution to the leptonic decay of the pion:
\begin{equation}
\Gamma_{\pi\to \ell \nu_\ell} = \case{1}{4\pi}\,f_\pi^2\,G_F^2\,|V_{ud}|^2\,
m_\pi\,m_\ell^2\,\left(1-m_\ell^2/m_\pi^2\right)^2\,,
\end{equation}
$G_F=1.166\times 10^{-5}\,$GeV$^{-2}$, $|V_{ud}|=0.975$, and
$m_e=0.511\,$MeV, $m_\mu= 0.106\,$GeV.

It is a model independent consequence of the axial-vector Ward-Takahashi
identity that\cite{mrt98}
\begin{equation}
\label{ggmor}
f_\pi\,m_\pi^2 = 2\,m_R(\zeta)\,r_\pi(\zeta)\,.
\end{equation}
In the chiral limit
\begin{equation}
\label{rpiqbq}
\lim_{\hat m \to 0}\,r_\pi(\zeta) = 
-\frac{1}{f_\pi^0}\langle \bar q q \rangle^0_\zeta\,,
\end{equation}
where $f_\pi^0$ is the chiral limit decay constant, and in this model
\begin{eqnarray}
\nonumber
\lefteqn{f_\pi^0=0.088\,{\rm GeV}\,,\; 
-\langle \bar q q \rangle^0_{1\,{\rm GeV}^2}= (0.235\,{\rm GeV})^3}\\
& \Rightarrow & 
r_\pi^0(1\, {\rm GeV}^2)=(0.384\,{\rm GeV})^2\,.
\end{eqnarray}

\subsection{Scalar Channel}
The analogue of Eq.~(\ref{psvtx}) for the $0^{++}$ vertex is presented in
Eq.~(\ref{scavtx}).  However, the combination of rainbow and ladder
truncations is not certain to provide a reliable approximation in the scalar
sector because here the cancellations described above do not
occur\cite{cdrqcII}.  This is entangled with the phenomenological
difficulties encountered in understanding the composition of scalar
resonances below $1.4\,$GeV\cite{mikescalars,mikescalar2,mikenew}.  For the
isoscalar-scalar vertex the problem is exacerbated by the presence of
timelike gluon exchange contributions to the kernel, which are the analogue
of those diagrams expected to generate the $\eta$-$\eta^\prime$ mass
splitting in BSE studies\cite{etareinhard}.  Nevertheless, in the absence of
an improved, phenomenologically efficacious kernel we employ
Eq.~(\ref{scavtx}) in the expectation that it will provide some qualitatively
reliable insight.  (This is justified {\it a posteriori}.)

The scalar functions in Eq.~(\ref{scavtx}) exhibit a simple pole at
$\vec{P}^2+m_{\sigma}^2 = 0$, Eqs.~(\ref{scpole},\ref{scnorm}), with residue
\begin{equation}
\label{rsc}
\delta^{\alpha\beta}\, r_{\sigma} = Z_4\,{\rm tr}\int_{l,q}^{\bar\Lambda}\,
\case{1}{2}\tau^\alpha\,\chi_{\sigma}^\beta(q_{\omega_l};\vec{P})\,,
\end{equation}
where $\chi^\alpha_{\sigma}$ is an obvious analogue of $\chi_{\rm \pi}^i$.
Since a $V-A$ current cannot connect a $0^{++}$ state to the vacuum the
scalar meson does not appear as a pole in the vector vertex; i.e, 
\begin{equation}
 \delta^{\alpha\beta}\, \vec{P}\,f_{\sigma}  = 
Z_2^A\,{\rm tr}\int_{l,q}^{\bar\Lambda}\,
\case{1}{2}\tau^\alpha\, \vec{\gamma}\,
\chi_\sigma^\beta(q_{\omega_l};\vec{P})\equiv 0\,.
\end{equation}
The homogeneous equation for the scalar bound state amplitude is obtained
from Eqs.~(\ref{scavtx},\ref{scpole}).  

\subsection{Two-body Decays}
\label{2body}
The $\sigma$ and $\pi$ bound state amplitudes along with the dressed-quark
propagators are necessary elements in the definition of the impulse
approximation to hadronic matrix elements.  For example, the
isoscalar-scalar-$\pi\pi$ coupling is described by the matrix element in
Eq.~(\ref{sigpipi}).  This yields the width
\begin{eqnarray}
\Gamma_{\sigma\to(\pi\pi)} & = &
\case{3}{2}\,g_{\sigma\pi\pi}^2\,\frac{\sqrt{1-4m_\pi^2/m_\sigma^2}}{16\,\pi\,
m_{\sigma}}\, ,
\end{eqnarray}
which vanishes if $m_\sigma< 2\,m_\pi$.  A contemporary analysis of $\pi\pi$
data identifies a $\bar u u$+$\bar d d$ scalar with\cite{mikescalar2}
\begin{equation}
m_\sigma \approx 0.46\,{\rm GeV}\,,\; \Gamma_\sigma\sim 0.22{\rm -}0.47\,{\rm
GeV} \,,
\end{equation}
which corresponds to $g_{\sigma \pi\pi}\sim 2.1$--$3.0\,$GeV
$=4.5$--$6.6\,m_\sigma$.

We also consider the electromagnetic decay of the neutral pion, for which the
dressed-quark-photon vertex is also required in calculating the impulse
approximation to the coupling.  Quantitatively reliable numerical solutions
of the $T=0$ vector vertex equation are now available\cite{Maris:1999}.
However, this anomalous coupling is insensitive to details and an accurate
result requires only that the dressed vertex satisfy the vector
Ward-Takahashi identity.  This is similar to the pion form factor: the
$q^2=0$ value is fixed by current conservation and only the charge radius
responds to changes in the vertex\protect\cite{Maris:1999}, and the
$\gamma^\ast \pi^0\to \gamma$ transition form factor whose value is fixed at
the real photon point\cite{cdrpion} but whose $q^2$-evolution is sensitive to
details of the vertex\cite{dubravko}.

An efficacious $T=0$ vertex {\it Ansatz} is given in
Eq.~(\ref{bcvtx0})\cite{bc80,adnan} and using this in calculating the $\hat m
=0$ $\pi^0\to \gamma\gamma$ coupling (Eq.~(\ref{pi0gg}) for $T\to 0$) one
obtains\cite{cdrpion,mrpion}
\begin{equation}
g^0_{\pi^0\gamma\gamma} = \case{1}{2}
\end{equation} 
{\it independent} of the model parameters.  Using this chiral limit coupling
on-shell
\begin{equation}
\Gamma_{\pi^0\to\gamma\gamma} \approx 
\frac{m_\pi^3}{16\pi}\,
\frac{\alpha_{\rm em}^2}{\pi^2}\,
\left(\frac{g^0_{\pi^0\gamma\gamma}}{f_\pi}\right)^2
= \frac{m_\pi^3}{64\pi}\,
\left(\frac{\alpha_{\rm em}}{\pi\,f_\pi}\right)^2\,;
\end{equation}
i.e., $7.7\,$keV cf. the experimental value\cite{pdg}: $7.7\pm 0.6$.

\section{Meson Properties}
\label{sec3}
Solving the (in)homogeneous BSE at $T=0$ is a demanding numerical task
because an accurate solution requires a large amount of computer memory
and/or time.  These problems are exacerbated at $T\neq 0$ because of the loss
of $O(4)$ invariance, which is manifest in the separation of the
four-momentum into a three-momentum and a Matsubara frequency.  Hence in the
sum over fermion Matsubara modes we usually limit ourselves to $l =
-4,\ldots,4$.  At the critical temperature this corresponds to
$\omega_{l=4}>4.0\,$GeV, which is greater than the other mass-scales in the
problem, and yields results that are numerically accurate to within $\sim
5$\%.  Reproducing the $T=0$ limit requires many more Matsubara
modes\cite{firstT}, which precludes that as a check of our numerical method.
Instead we estimate the error by reducing the number of modes and comparing
the results.  Our discretisation of the three-momentum grids in the quark DSE
and BSE is less seriously limited, and we typically employ $\gsim 1000$
Gaussian quadrature points.

\subsection{Chiral limit}
\label{sub3a}
For $\hat m =0$ the $T=0$ analogues of the inhomogeneous BSEs,
Eqs.~(\ref{psvtx},\ref{scavtx}), exhibit poles at
\begin{equation}
m_\pi = 0\,\;\; {\rm and} \;\;m_\sigma= 0.56\,{\rm GeV}\,,
\end{equation}
with $m_\sigma = 0.59\,$GeV at $\hat m = 5.7\,$MeV.  A low-mass scalar is
typical of the rainbow-ladder truncation.  However, there is some model
sensitivity; e.g., cf. this result with $m_\sigma = 0.59\,$GeV in
Ref.~\cite{herman2}, $m_\sigma = 0.67\,$GeV using the model of
Ref.~\cite{mt99} and $m_\sigma = 0.72\,$GeV in the separable model of
Ref.~\cite{conrad}.  These four independent calculations give an
average-$m_\sigma= 0.64\,$GeV with a standard deviation of $10\,$\%.  The
rainbow-ladder truncation yields degenerate isoscalar and isovector bound
states, and ideal flavour mixing in the $3$-flavour case.  Hence the
distribution of mass estimates between that of the isoscalar $\sigma$ and
isovector $a_0(980)$ might be anticipated.  In contrast the three comparison
studies, Refs.~\cite{herman2,mt99,conrad}, give $m_\omega= m_\rho =
0.75\,$GeV with a standard deviation of $<2$\%, illustrating the
dependability of the truncation in the vector channel.

As already remarked, improvements to the kernel are required in the scalar
channel.  In the isoscalar-scalar channel, because $\Gamma_\sigma/m_\sigma$
is large, it may even be necessary to include couplings to the dominant
$\pi\pi$ mode, which can be handled perturbatively in the $\omega$-$\rho$
sector\cite{mikenew,rhomass}.  In the absence of such corrections, the
$\sigma$ properties elucidated herein are strictly only those of an idealised
chiral partner of the $\pi$.  Hitherto no model bound-state description
escapes this caveat.

\begin{figure}[t] 
\centering{\
\epsfig{figure=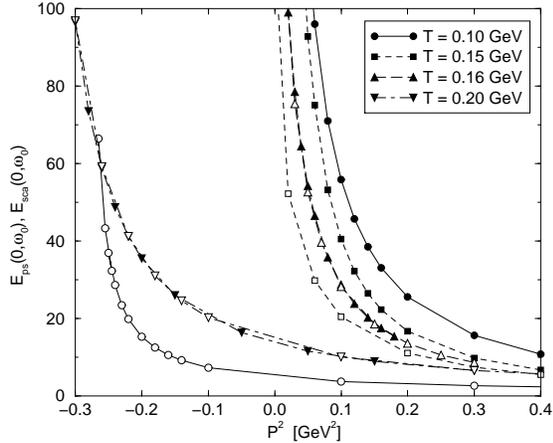,height=6.0cm}}\vspace*{0.5\baselineskip} 

\caption{Leading Dirac amplitude for the pseudoscalar (shaded symbols) and
scalar (open symbols) vertices, $E$ in
Eqs.~(\protect\ref{psvtxform},\protect\ref{scvtxform}), evaluated at
$(\vec{p}=0,\omega_0)$ and plotted as function of $\vec{P}^2$ in the chiral
limit; i.e., $E(\vec{p}=0,\omega_0;\vec{P}^2)$, for a range of temperature
values.  The bound state poles are evident in each case.  (We use
$\zeta=19\,$GeV throughout.)\label{ihbse}}
\end{figure}

The evolution with $T$ of the pole positions in the solution of the
inhomogeneous BSEs is illustrated in Fig.~\ref{ihbse}, from which it is clear
that: 1) at the critical temperature, $T_c$, we have degenerate, massless
pseudoscalar and scalar bound states; and 2) the bound states persist above
$T_c$, becoming increasingly massive with increasing $T$.  These features are
also observed in numerical simulations of lattice-QCD\cite{echaya2}.

\begin{figure}
\centering{\
\epsfig{figure=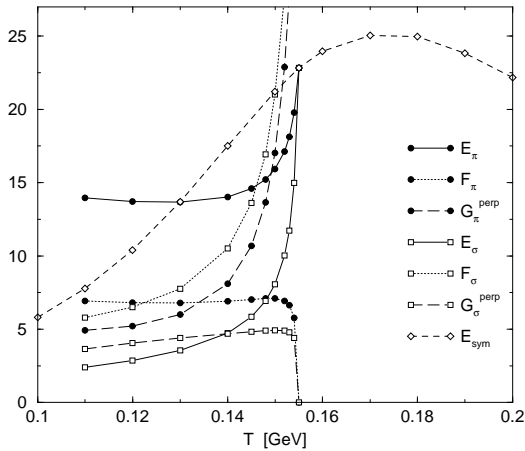,height=6.0cm}}\vspace*{\baselineskip} 

\caption{\label{bsamps0}Dirac amplitudes characterising the $\hat m =0$
pseudoscalar and scalar bound states, evaluated at $(\vec{p}=0,\omega_0)$.
$G_\pi^\perp$, $G_\pi^\|$ and $F_\sigma$ behave similarly: as $T\to T_c^-$
their value at $(\vec{p}=0,\omega_0)$ increases rapidly while the domain on
which the functions are nonzero shrinks quickly.  $E_{\rm sym}$ is the
amplitude calculated in the chirally symmetric $B_0\equiv 0$ phase: above
$T_c$, $E_\pi=E_\sigma = E_{\rm sym}$ while all other amplitudes vanish.}
\end{figure}

We obtain the bound state amplitudes from the homogeneous Bethe-Salpeter
equations, which Fig.~\ref{ihbse} demonstrates are certain to have a
solution.  Their $T$-evolution is depicted in Fig.~\ref{bsamps0}, which
indicates that: 1) in both cases all but the leading Dirac amplitude vanishes
above $T_c$; and 2) the surviving pseudoscalar amplitude is pointwise
identical to the surviving scalar one.  These results indicate that the
chiral partners are {\it locally} identical above $T_c$, they do not just
have the same mass.  

It is easy to understand this algebraically.  The BSE is a set of coupled
homogeneous equations for the Dirac amplitudes.  Below $T_c$ each of the
equations for the subleading Dirac amplitudes has an ``inhomogeneity'' whose
magnitude is determined by $B_0$, the scalar piece of the quark self energy
which is dynamically generated in the chiral limit.  $B_0$ vanishes above
$T_c$ eliminating the inhomogeneity and allowing a trivial, identically zero
solution for each of these amplitudes.  Additionally, with $B_0\equiv 0$ the
kernels in the equations for the dominant pseudoscalar and scalar amplitudes
are identical, and hence so are the solutions.

It follows from these results that the Goldberger-Treiman-like
relation\cite{mrt98}
\begin{equation}
\label{gtlrel}
f_\pi^0 E_\pi(p_{\omega_k};0) = B_0(p_{\omega_k})\,,
\end{equation}
is satisfied for all $T$ only because both $f_\pi^0$ and $B_0(p_{\omega_k})$
are equivalent order parameters for chiral symmetry restoration.  This
possibility was overlooked in Ref.~\cite{firstT}.  Further, above $T_c$, the
other constraints on the chiral-limit pion Bethe-Salpeter amplitude derived
in Ref.~\cite{mrt98} from the axial-vector Ward-Takahashi identity are
trivially satisfied.

\begin{figure}[t]
\centering{\
\epsfig{figure=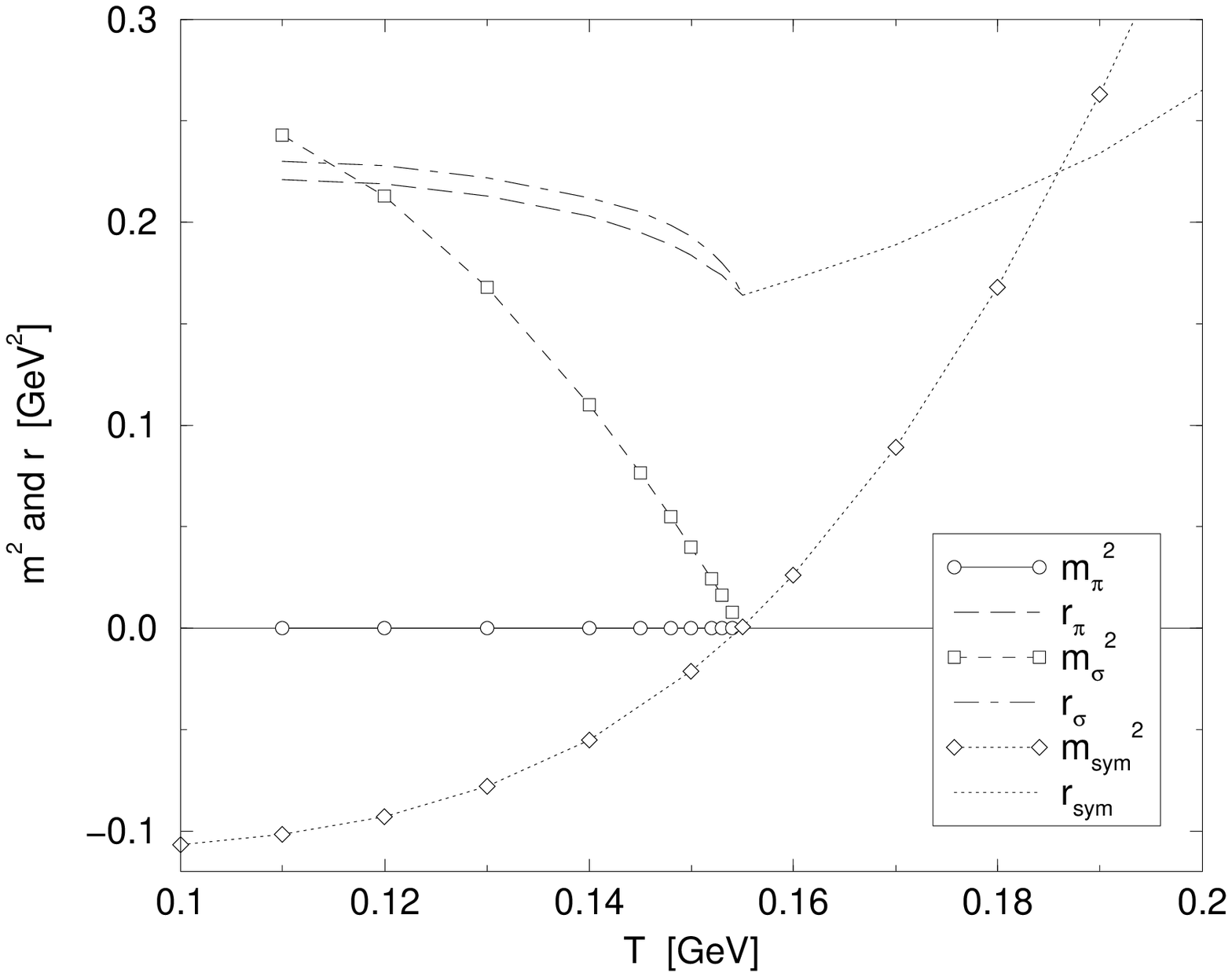,height=6.0cm}}\vspace*{0.5\baselineskip}

\centering{\
\epsfig{figure=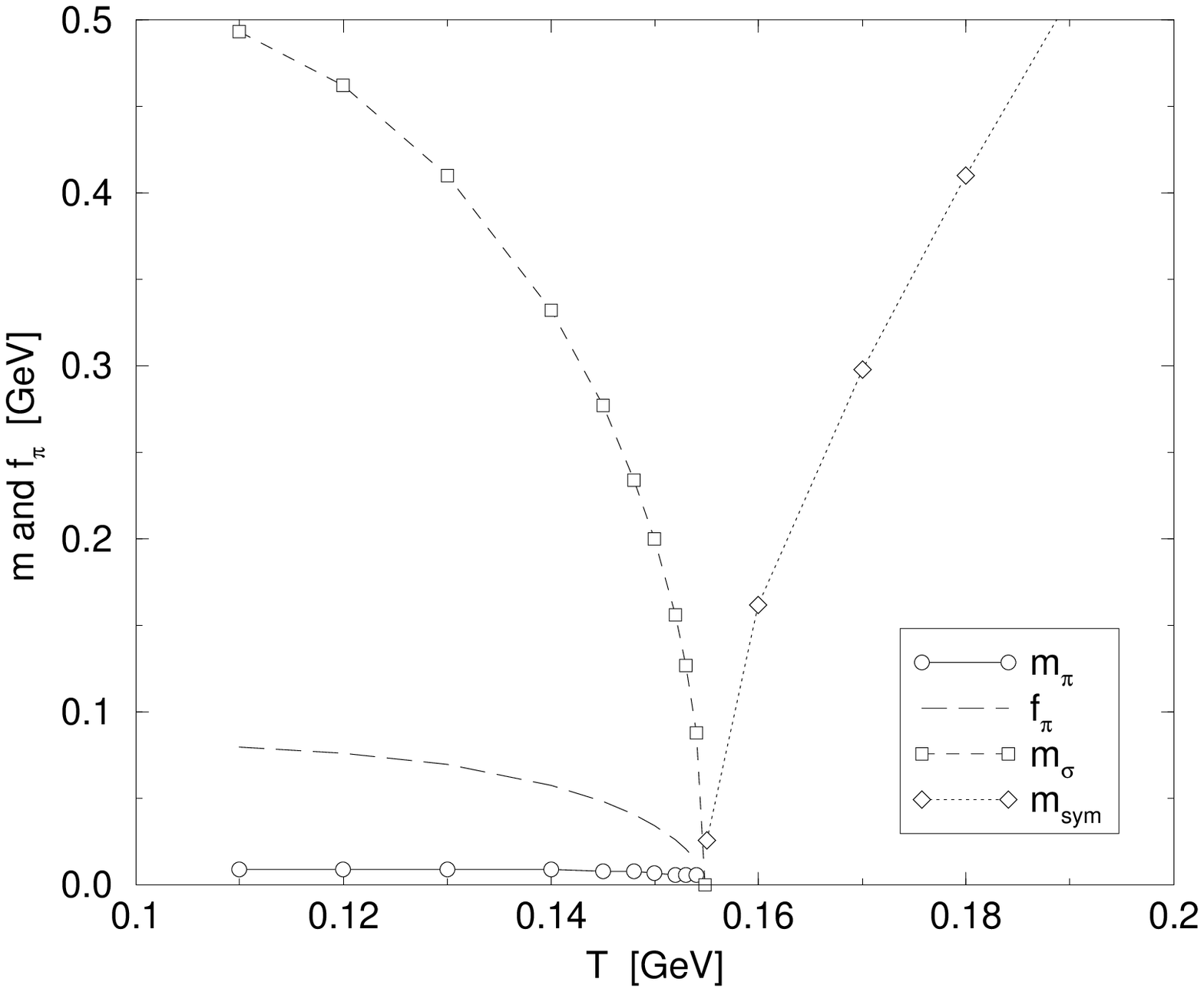,height=6.0cm}}\vspace*{0.5\baselineskip}

\caption{\label{obs0}$\hat m = 0$ results.  {\it Upper panel}: $T$-dependence
of the meson masses-squared and pole-residue matrix elements,
Eqs.~(\protect\ref{rpi},\protect\ref{fpi},\protect\ref{rsc}), along with the
mass-squared: $m_{\rm sym}^2$, and residue: $r_{\rm sym}$, calculated in the
chirally symmetric $B_0\equiv 0$ phase.  For $T\leq T_c$, $m_{\rm sym}^2<0$,
which is a signal of the instability of the chirally symmetric phase at low
$T$.  For $T>T_c$, $m_\pi^2=m_\sigma^2=m_{\rm sym}^2$.  {\it Lower panel}:
$T$-dependence of the masses and pion decay constant.  $m_\pi=0$ within
numerical error.}
\end{figure}

Using the bound state amplitudes and dressed-quark propagators we calculate
the matrix elements discussed in Sec.~\ref{sec2}.  Their chiral limit
$T$-dependence is depicted in Fig.~\ref{obs0}.  As indicated by
Fig.~\ref{ihbse}, the pseudoscalar and scalar bound states are massive and
degenerate above $T_c$.

Below $T_c$ the scalar meson residue in the scalar vertex, $r_\sigma$ in
Eq.~(\ref{rsc}), is a little larger than the residue of the pseudoscalar
meson in the pseudoscalar vertex, $r_\pi$ in Eq.~(\ref{rpi}).  However, they
are nonzero and equal above $T_c$, which is an algebraic consequence of $B_0
\equiv 0$ and the vanishing of the subleading Dirac amplitudes.  
As a {\it bona fide} order parameter for chiral symmetry restoration
\begin{equation}
f_\pi \propto (1-T/T_c)^{\beta}\,,\; T/T_c\lsim 1\,,
\end{equation}
where $\beta$ is the zero-external-field critical-exponent for chiral
symmetry restoration ($\beta = 1/2$ in rainbow-ladder
models)\cite{Holl:1999}.  $f_\pi = 0$ and $r_\pi \neq 0$ for $T> T_c$
demonstrates that the pion disappears as a pole in the axial-vector
vertex\cite{firstT} but persists as a pole in the pseudoscalar vertex.  

Our analysis yields
\begin{equation}
m_\sigma \propto (1-T/T_c)^{\beta}\,,\; T/T_c\lsim 1\,, 
\end{equation}
within numerical errors, which is evident in Fig.~\ref{obs0}: $m_\sigma^2$
follows a linear trajectory in the vicinity of $T_c$.  Such behaviour in the
isoscalar-scalar channel might be anticipated because this channel has vacuum
quantum numbers and hence the bound state is a strong interaction analogue of
the electroweak Higgs boson\cite{mikenew}.

$m_{\rm sym}^2$ in Fig.~\ref{obs0} is the mass obtained when the chirally
symmetric solution of the quark DSE is used in the BSE.  ($B_0 \equiv 0$ is
always a solution in the chiral limit.)  For $T>T_c$, $m_{\rm sym}^2(T)$ is
the unique meson mass-squared trajectory.  However, for $T<T_c$, $m_{\rm
sym}^2<0$; i.e., the solution of the BSE in the Wigner-Weyl phase exhibits a
tachyonic solution (cf. the Nambu-Goldstone phase masses: $m_\sigma^2 >
m_\pi^2 = 0$).  By analogy with the $\sigma$-model this tachyonic mass
indicates the instability of the Wigner-Weyl phase below $T_c$.  It
translates into the statement that the pressure is not maximal in this phase.

\begin{figure}[t]
\centering{\
\epsfig{figure=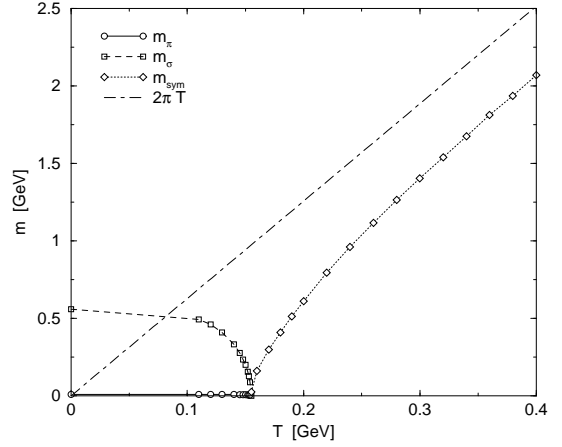,height=6.0cm}}\vspace*{0.5\baselineskip} 

\caption{\label{screen}$T$-dependence of for large $T$ with $\hat m=0$, see
Fig.~\protect\ref{obs0}.  $m_\pi = m_\sigma$ for $T>T_c$ and $m/(2\pi T) \to
1^-$.  This behaviour persists with $\hat m \neq 0$, however, providing an
illustration requires significantly more computer time.  (That is easier in a
simpler
model~\protect\cite{Blaschke:1999,peterNew}.)}\vspace*{0.5\baselineskip}
\end{figure}

Figure~\ref{screen} depicts the evolution of the (common) meson mass at large
$T$.  As expected in a gas of weakly interacting quarks and gluons
\begin{equation}
\label{screeneq}
\frac{m_{\rm meson}}{2\, \omega_0} \to 1^-\,,
\end{equation}
where $\omega_0 = \pi T$ is a quark's zeroth Matsubara frequency and
``screening mass.''  

\subsection{Nonzero light current-quark masses}
\begin{figure}
\centering{\
\epsfig{figure=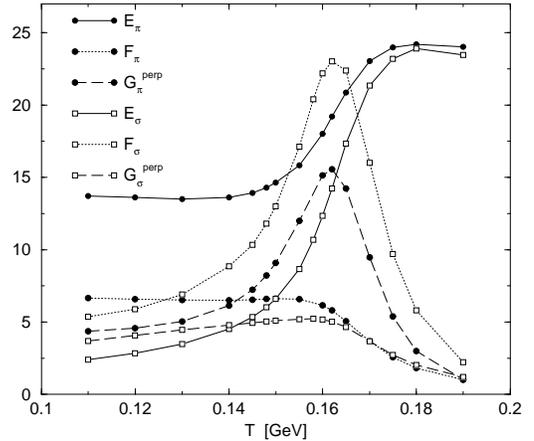,height=6.0cm}}\vspace*{0.5\baselineskip}  

\caption{\label{bsam}Dirac amplitudes characterising the $\hat m \neq 0$
pseudoscalar and scalar bound states, evaluated at $(\vec{p}=0,\omega_0)$ and
plotted as a function of temperature.}
\end{figure}

It is straightforward to repeat the calculations of Sec.~\ref{sub3a} for
nonzero current-quark masses where chiral symmetry restoration with
increasing $T$ is exhibited as a crossover rather than a phase transition.
The solutions of the inhomogeneous BSEs again exhibit a pole for all $T$ and
we determine the bound state amplitudes from the associated homogeneous
equations.  Their $T$-dependence is characterised in Fig.~\ref{bsam}, which
shows that for $\hat m \neq 0$ they are locally identical for $T \gsim
\case{4}{3} T_c$.

\begin{figure}[h]
\centering{\
\epsfig{figure=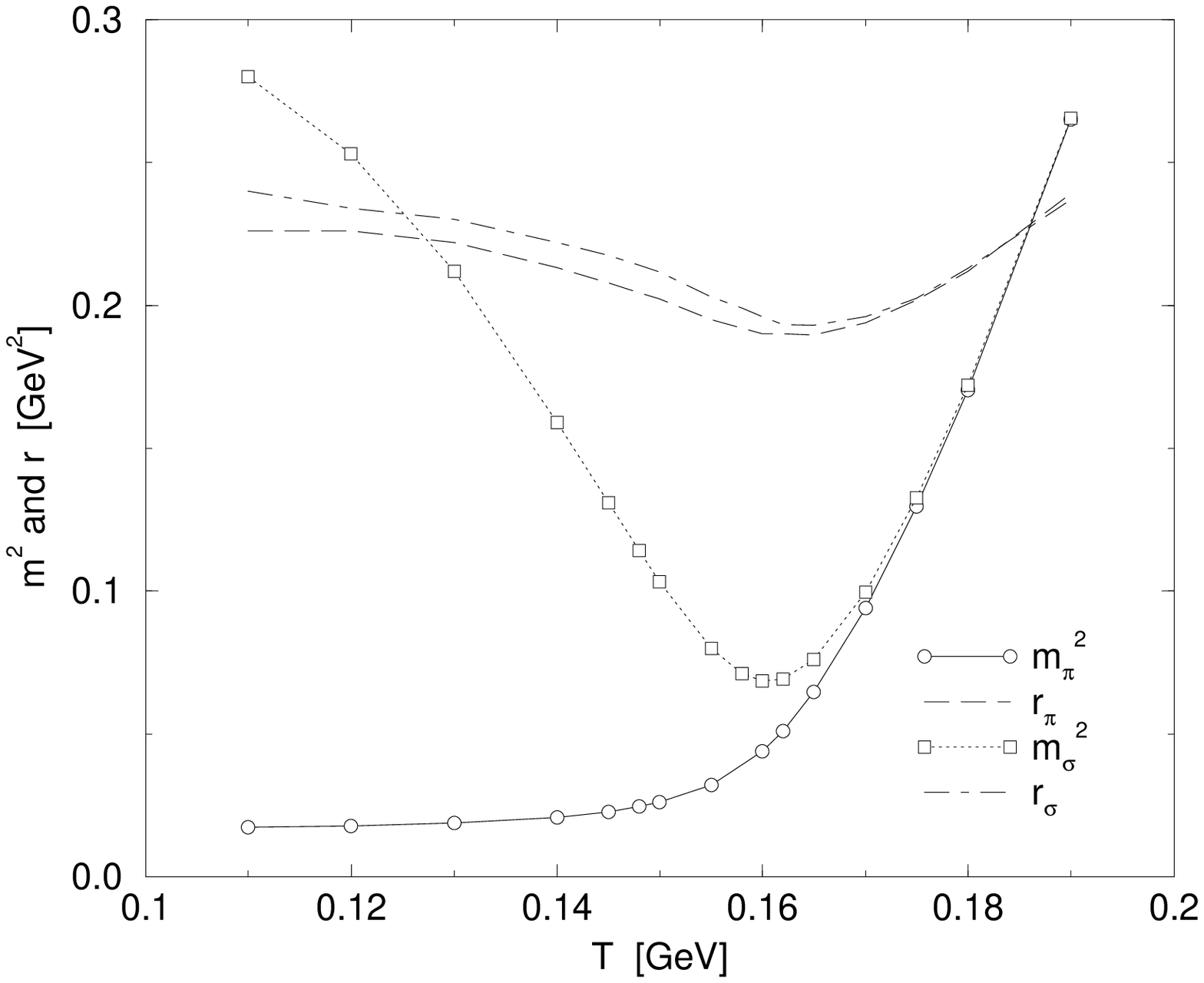,height=6.0cm}}\vspace*{0.5\baselineskip} 

\centering{\
\epsfig{figure=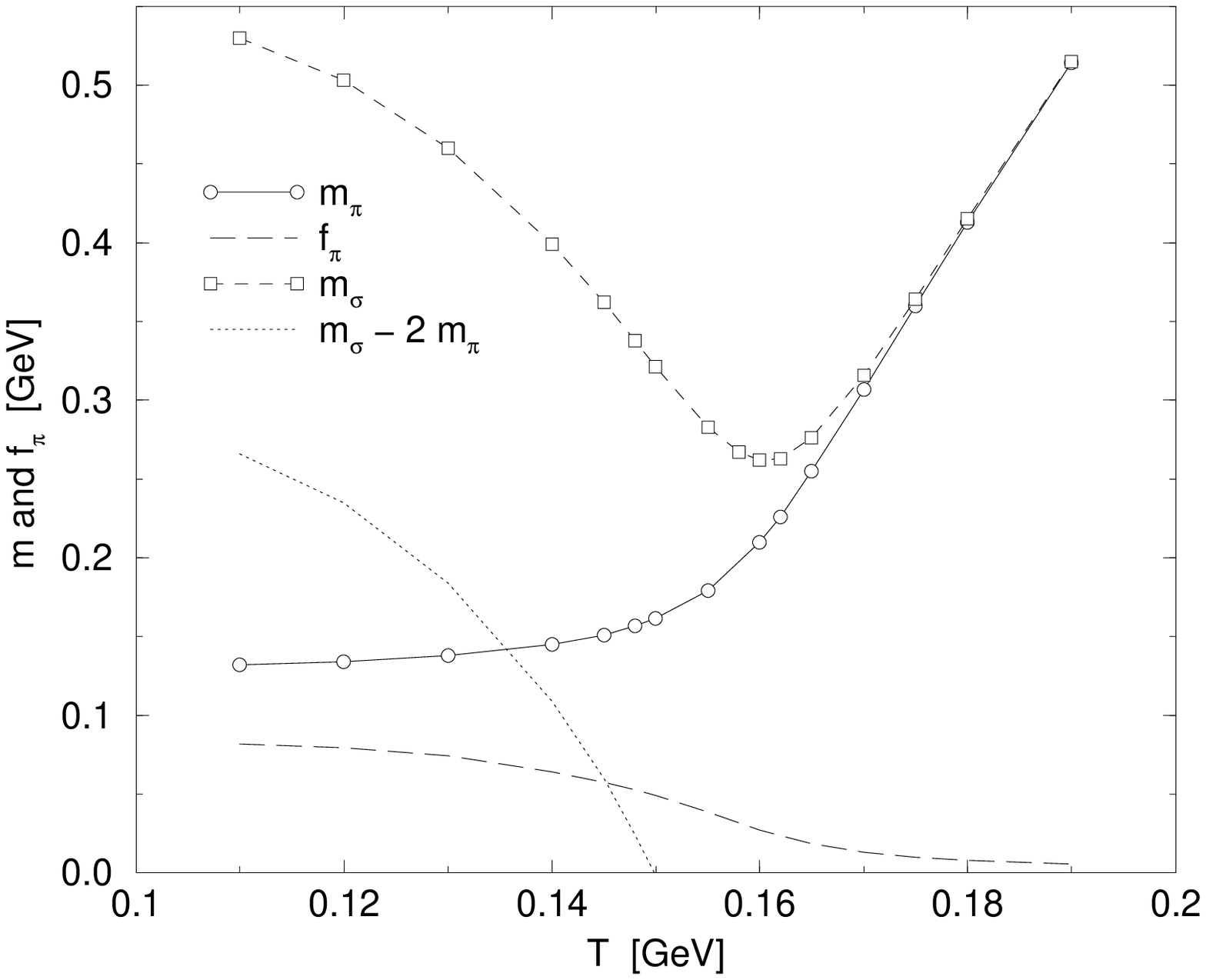,height=6.0cm}}\vspace*{0.5\baselineskip} 

\caption{\label{obsm}$T$-dependence of the meson masses and pole-residue
matrix elements,
Eqs.~(\protect\ref{rpi},\protect\ref{fpi},\protect\ref{rsc}), for $\hat m
\neq 0$. }
\end{figure}

Figure~\ref{obsm} is the $\hat m\neq 0$ analogue of Fig.~\ref{obs0}.  That
the transition has become a crossover is evident in the behaviour of $f_\pi$.
The meson masses become indistinguishable at $T\sim 1.2 \,T_c$, a little
before the local equivalence is manifest in Fig.~\ref{bsam}, which is
unsurprising given that the mass is an integrated quantity.  The small
difference between $r_\sigma$ and $r_\pi$ below $T_c$ is again evident and
they assume a common value at the same temperature as the masses.

\begin{figure}
\centering{\
\epsfig{figure=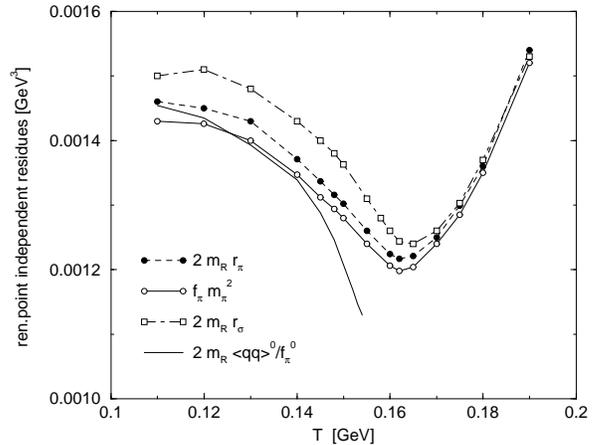,height=6.0cm}}\vspace*{0.5\baselineskip}  

\caption{\label{avwtipic}$T$-dependence of the residue of the pion pole in
the axial-vector vertex, $f_\pi m_\pi^2$, and in the pseudoscalar vertex,
$2\,m_R(\zeta)\, r_\pi(\zeta)$.  They are equal within numerical errors as
required by the axial-vector Ward-Takahashi identity.  For comparison we also
plot $2\,m_R(\zeta)\, r_\pi^0(\zeta)$ and the residue of the scalar meson
pole in the scalar vertex, $2\,m_R(\zeta) \,r_\sigma(\zeta)$.}
\end{figure}

Figure~\ref{avwtipic} illustrates the preservation of the axial-vector
Ward-Takahashi identity via the mass formula of Eq.~(\ref{ggmor}).  The
magnitude and $T$-dependence of both sides are equal within numerical errors
above {\it and} below the crossover.  $m_R(\zeta) \,r_\pi^0(\zeta)$ is the
renormalisation point independent quantity that appears in the pion's current
algebra mass formula: $m_R(\zeta)\, r_\pi^0(\zeta)$ and $m_R(\zeta)
\,r_\pi(\zeta)$ differ by $< 5$\% until $T>0.95\,T_c$.  This comparison
illustrates the $T$-domain on which the current algebra formula is valid and
the analysis of Ref.~\cite{yura2}.  It was noted in Ref.~\cite{Holl:1999}
that
\begin{equation}
r_\pi^0(\zeta)/f_\pi^0 \propto (1-T/T_c)^{-\beta}\,,\; T/T_c\lsim 1\,,
\end{equation}
which is qualitatively apparent from Figs.~\ref{obs0} and \ref{avwtipic}.

\subsection{Triangle Diagrams}
We also studied the matrix elements describing the two-body decays discussed
in Sec.~\ref{2body}.  With every element in the calculation only known
numerically this too is a challenging numerical exercise, which we simplified
by a judicious choice of the external momenta.

\begin{figure}
\centering{\
\epsfig{figure=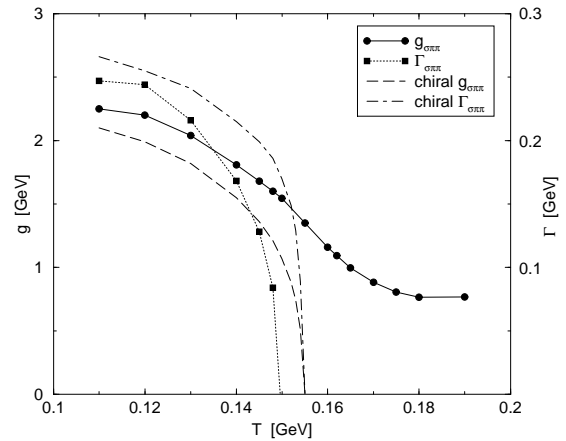,height=6.0cm}}\vspace*{0.5\baselineskip} 

\caption{\label{widtha}$T$-dependence of the iso\-scalar-scalar-$\pi\pi$
coupling and width, both in the chiral limit and for realistic light
current-quark masses. The phase space factor $(1-(2m_\pi/m_\sigma)^2)^{1/2}$
is $\theta(T_c-T)$ in the chiral limit but nontrivial for $\hat m \neq 0$,
vanishing at $T\approx 0.98 \,T_c$; i.e., this decay channel closes at a
temperature just $2$\% less-than $T_c$.}
\end{figure}

The chiral limit isoscalar-scalar-$\pi\pi$ coupling and width obtained from
Eq.~(\ref{sigpipi}) are depicted in Fig.~\ref{widtha}, which indicates that
both vanish at $T_c$ in the chiral limit.  Again this can be traced to $B_0
\to 0$.  For $\hat m \neq 0$, the coupling reflects the crossover.  However,
that is moot because the width vanishes just below $T_c$ where the
isoscalar-scalar meson mass falls below $2 m_\pi$ and the phase space factor
vanishes (see the lower panel of Fig.~\ref{obsm}).

\begin{figure}
\centering{\
\epsfig{figure=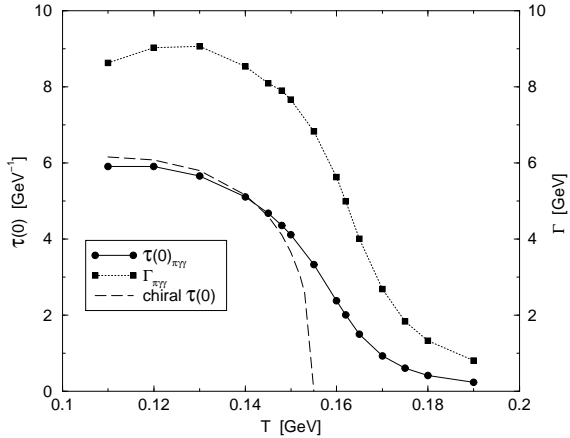,height=6.0cm}}\vspace*{0.5\baselineskip}

\caption{\label{widthb}$T$-dependence of the coupling ${\cal T}(0)$ in
Eq.~(\protect\ref{anomT}) and the $\pi^0\to \gamma\gamma$ width.}
\end{figure}

\begin{figure}
\centering{\
\epsfig{figure=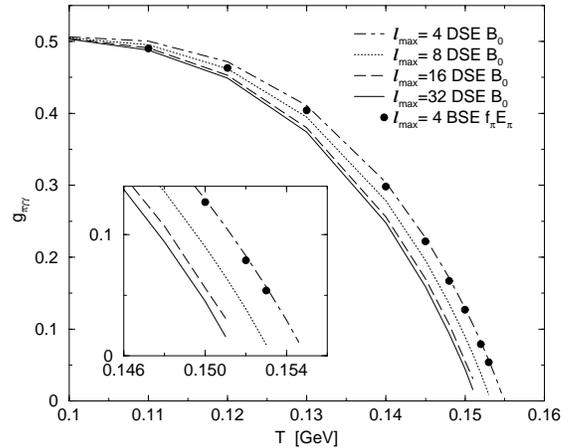,height=6.0cm}}\vspace*{0.5\baselineskip}

\caption{\label{comppigg}$T$-dependence of $g^0_{\pi\gamma\gamma}= f_\pi^0
{\cal T}(0)$.  Within numerical error, $g^0_{\pi\gamma\gamma} \propto
(1-T/T_c)$.  Near $T_c$, $f_\pi \propto (1-T/T_c)^{1/2}$ because our model
has mean field critical exponents and hence ${\cal T}(0)\propto
(1-T/T_c)^{1/2}$.}
\end{figure}

The $T$-dependence of the $\pi^0\gamma\gamma$ coupling, which saturates the
Abelian anomaly at $T=0$, is calculated from
Eqs.~(\ref{pigganom},\ref{anomT}) and depicted along with the width in
Fig.~\ref{widthb}.  In the chiral limit the width is identically zero because
$m_\pi=0$ and the interesting quantity is: ${\cal
T}(0)=g^0_{\pi^0\gamma\gamma}/f_\pi^0$.  Clear in the figure is that ${\cal
T}(0)$ vanishes at $T_c$.  It vanishes with a mean field critical exponent,
as is most easily inferred from Fig.~\ref{comppigg}.  (An accurate
calculation is possible because Eq.~(\ref{gtlrel}) obviates the need for a
solution of the BSE.)  Thus, in the chiral limit, the coupling to the
dominant decay channel closes for both charged {\it and} neutral pions.
These features were anticipated in Ref.~\cite{pisarski}.  Further, as is
evident in Fig.~\ref{comppigg}, our calculated ${\cal T}(0)$ is monotonically
decreasing with $T$, supporting the perturbative O($T^2/f_\pi^2)$ analysis in
Ref.~\cite{pisarski2}.

For $\hat m \neq 0$ both the coupling: $g_{\pi^0\gamma\gamma}/f_\pi$, and the
width exhibit the crossover with a slight enhancement in the width as $T\to
T_c$ due to the increase in $m_\pi$.  There are similarities between these
results and those of Ref.~\cite{yura} although the $T$-dependence herein is
much weaker because our pion mass approaches twice the $T \neq 0$ free-quark
screening-mass from below, never reaching it, Eq.~(\ref{screeneq}); i.e., the
continuum threshold is not crossed.

\section{Summary and Conclusion}
\label{sec4}
We employed a renormalisation group improved rainbow-ladder truncation of the
quark Dyson-Schwinger equation, and pseudoscalar and scalar Bethe-Salpeter
equations to estimate the $T$-dependence of a range of properties that
characterise correlations in these channels.  The rainbow-ladder truncation
is quantitatively reliable in the pseudoscalar channel.  However, that is not
certain in the scalar channel where the cancellations that assist in the
pseudoscalar channel are not apparent.  Nevertheless, we anticipate that many
of the features we exposed in the scalar channel are qualitatively correct.

The solutions of the inhomogeneous Bethe-Salpeter equation (BSE) exhibit
poles at all $T$, both above and below the critical temperature, and in the
chiral limit and for realistic light current-quark masses.  We use the
associated homogeneous BSEs to determine the masses, which correspond to the
screening masses determined in simulations of lattice-QCD, and find
$m_{\sigma,\pi} \to 2\pi T$ as $T\to \infty$.  

The BSE solutions indicate a {\it local} equivalence between the
isovector-scalar and -pseudoscalar correlations above the chiral restoration
transition/crossover;\footnote{We refrain from asserting this of the
isoscalar-scalar correlation because we cannot anticipate the effect of
timelike gluon exchange contributions present in the BSE kernel in this
channel.  However, they are kindred to those encountered in analysing the
realisation of $U_A(1)$ symmetry above the chiral
transition\protect\cite{echaya2,hirschegg97,etareinhard}.}
i.e., the scalar functions characterising the bound state amplitudes are
identical, and from this follows equality of the masses and many of the
matrix elements.\footnote{The evolution to equality of the masses has been
observed in numerical simulations of lattice-QCD and in the exploration of
other models that accurately describe dynamical chiral symmetry breaking.}
Further, the axial-vector Ward-Takahashi identity and the pseudoscalar mass
formula that is its corollary are valid both above and below the crossover.

For realistic light current-quark masses the isoscalar-scalar meson mass does
not fall below $2 m_\pi$ until very near the transition temperature.  Hence
this dominant decay channel remains open almost until the phase boundary is
crossed.  Once it is crossed, however, only electroweak decay channels are
open.  In the chiral limit the anomalous two-photon coupling constant
vanishes at $T_c$ just as does $f_\pi$, the coupling that determines the
strength of the leptonic charged pion decay.  For realistic masses, however,
the widths for the leptonic charged pion mode and two-photon $\pi^0$ mode
remain significant in the vicinity of the crossover.  

The construction of a DSE-BSE truncation that allows for an improved
description of the scalar channel at $T=0$ would provide the foundation for a
significant improvement of our analysis.  Employing an {\it Ansatz} for the
kernel of the quark DSE whose infrared form exhibits some $T$-dependence,
perhaps constrained by lattice simulations of the string
tension\cite{echaya2}, may also be interesting.  However, given the results
of Refs.~\cite{Holl:1999,reinhard}, we do not expect such a modification to
have a significant qualitative impact.  In the absence of such improvements
we nevertheless expect the local equivalence we have elucidated to be
exhibited by all isovector chiral partners in the strong interaction
spectrum.  However, the explicit demonstration of this is difficult; e.g., in
the $\rho$-$a_1$ complex the bound state amplitudes have {\it eight}
independent amplitudes even at $T=0$ compared with the four in the
pseudoscalar and scalar amplitudes at $T\neq 0$.

\section*{Acknowledgments}
We acknowledge interactions with D. Blaschke and Yu.L. Kalinovsky.  C.D.R. is
grateful for the support and hospitality of the Special Centre for the
Subatomic Structure of Matter at the University of Adelaide during a visit in
which some of this work was conducted, and both C.D.R. and S.M.S. are
grateful for the same from the Physics Department at the University of
Rostock during a joint visit in which aspects of this work were completed.
S.M.S. acknowledges financial support from the A.v.~Humboldt foundation.
This work was supported by the US Department of Energy, Nuclear Physics
Division, under contract no. W-31-109-ENG-38, the National Science Foundation
under grant nos. INT-9603385 and PHY97-22429, and benefited from the
resources of the National Energy Research Scientific Computing Center.

\appendix
\section*{Collected Formulae}
\subsection{Scalar Vertex}
The inhomogeneous ladder-like Bethe-Salpeter equation for the zeroth
Matsubara mode of the $0^{++}$ vertex is
\begin{eqnarray}
\nonumber
\lefteqn{
\Gamma_{\rm s}^\alpha(p_{\omega_k};P_0;\zeta)  =  Z_4
\,\case{1}{2}\tau^\alpha\mbox{\large\boldmath $1$}} \\
\nonumber 
\lefteqn{- \int_{l,q}^{\bar\Lambda} \,\case{4}{3}\,
g^2 D_{\mu\nu}(p_{\omega_k}-q_{\omega_l})} \\ 
\label{scavtx}
&  \times & \gamma_\mu S(q_{\omega_l}^+)\,\Gamma_{\rm
s}^i(q_{\omega_l};P_{\Omega_n};\zeta)\, S(q_{\omega_l}^-)\,\gamma_\nu\,,
\end{eqnarray}
where $\alpha= 0,1,2,3$ with $\tau^0 = {\rm diag}(1,1)$.  (NB: In this
truncation the isoscalar and isovector states are degenerate, which
exemplifies our observation that the ladder-like truncation is accurate for
vector mesons but requires improvement before it is quantitatively reliable
in the $0^{++}$ sector.) The solution has the form
\begin{eqnarray}
\nonumber
\lefteqn{\Gamma_{\rm s}^i(p_{\omega_k};\vec{P}) = 
\case{1}{2}\tau^\alpha\,\mbox{\large\boldmath $1$}\,
\left[ E_{\rm s} (p_{\omega_k};\vec{P})
+ i \vec{\gamma}\cdot\vec{p}\,G^{\|}_{\rm s} (p_{\omega_k};\vec{P})
\right.} \\
& +& 
\left.  
i \gamma_4\omega_k\,G^\perp_{\rm s} (p_{\omega_k};\vec{P})
+ i \vec{\gamma}\cdot\vec{P}\,\vec{p}\cdot\vec{P}\,
F_{\rm s} (p_{\omega_k};\vec{P})
\right]\,.
\label{scvtxform}
\end{eqnarray}
(NB: Here the requirement that the neutral mesons be charge conjugation
eigenstates shifts the $\vec{p}\cdot\vec{P}$ term cf. the $0^{-+}$
amplitude.)

The scalar functions in Eq.(\ref{scavtx}) exhibit a simple pole at
$\vec{P}^2+m_{\sigma}^2 = 0$:
\begin{equation}
\label{scpole}
\Gamma_{\rm s}^i(p_{\omega_k};\vec{P}) = 
\frac{r_{\sigma}(\zeta)}{\vec{P}^2 + m_{\sigma}^2}\,
\Gamma_{\sigma}^i(p_{\omega_k};\vec{P}) + {\rm regular}\,,
\end{equation}
where $\Gamma_{\sigma}^i(p_{\omega_k};\vec{P})$ is the canonically
normalised, $0^{++}$ bound state Bethe-Salpeter amplitude:
\begin{eqnarray}
\nonumber
\lefteqn{2 \delta^{\alpha \beta} \vec{P}  =  
{\rm tr}\,\int_{l,q}^{\bar\Lambda}\,\left\{
\Gamma_{\sigma}^\alpha(q_{\omega_l};-\vec{P}) \, 
\frac{\partial S(q_{\omega_l}^+)}{\partial \vec{P}} 
\Gamma_{\sigma}^\beta(q_{\omega_l};\vec{P}) \,  S(q_{\omega_l}^-)\right.}\\
\label{scnorm}
&& +
\left. \left.
\Gamma_{\sigma}^\alpha(q_{\omega_l};-\vec{P}) \, S(q_{\omega_l}^+)
\Gamma_{\sigma}^\beta(q_{\omega_l};\vec{P}) \,
\frac{\partial S(q_{\omega_l}^-)}{\partial \vec{P}} 
\right\}\right|_{\vec{P}^2= -m_{\rm s}^2}\!\!\!.
\end{eqnarray}

The dressed-quark propagator and canonically normalised Bethe-Salpeter
amplitudes make possible the definition of the impulse approximation to the
isoscalar-scalar-$\pi\pi$ matrix element: $\vec{p_1}^2=-m_\pi^2=\vec{p_2}^2$,
$(\vec{p}=\vec{p_1}+\vec{p_2})^2=-m_\sigma^2$,
\begin{eqnarray}
\nonumber \lefteqn{g_{\sigma\pi\pi}  := 
\langle \pi(\vec{p_1})\pi(\vec{p_2}) | \sigma(\vec{p})\rangle = }\\ &&
\nonumber 2 N_c{\rm tr}_D\int_{l,q}^{\bar\Lambda}\,
\Gamma_\sigma(k_{\omega_l};\vec{p})\,S_u(k_{++}) \\ && \times
i\Gamma_\pi(k_{0+};-\vec{p_1})\,
S_u(k_{+-})\,i\Gamma_\pi(k_{-0};-\vec{p_2})\,S_u(k_{--})\,,
\label{sigpipi}
\end{eqnarray}
$k_{\alpha\beta}= k_{\omega_l}+(\alpha/2)\vec{p_1}+(\beta/2)\vec{p_2}$, with
only the trace over Dirac indices remaining.

\subsection{Neutral Pion Decay}
An efficacious {\it Ansatz} for the dressed-quark-photon coupling at $T=0$
is\cite{bc80,adnan}:
\begin{eqnarray}
\label{bcvtx0}\lefteqn{i\Gamma_\mu(\ell_1,\ell_2) = 
i\Sigma_A(\ell_1^2,\ell_2^2)\,\gamma_\mu }\\ 
& & \nonumber 
+
(\ell_1+\ell_2)_\mu\,\left[\sfrac{1}{2}i\gamma\cdot (\ell_1+\ell_2) \,
\Delta_A(\ell_1^2,\ell_2^2) + \Delta_B(\ell_1^2,\ell_2^2)\right]\,;\\
&&  \Sigma_F(\ell_1^2,\ell_2^2) = \sfrac{1}{2}\,[F(\ell_1^2)+F(\ell_2^2)]\,,\\
&& \Delta_F(\ell_1^2,\ell_2^2) =
\frac{F(\ell_1^2)-F(\ell_2^2)}{\ell_1^2-\ell_2^2}\,,
\end{eqnarray}
where $F= A, B$; i.e., the scalar functions in the dressed-quark propagator:
$S^{-1}(p)= i \gamma\cdot p A(p^2) + B(p^2)$, so that this model is
completely determined by $S(p)$.  Improvements of this {\it Ansatz}, such as
those canvassed in Ref.~\cite{adnan}, do not have a qualitatively significant
effect in the present context.

The dressed-quark-photon vertex is a necessary element in the calculation of
the impulse approximation to the $\pi^0\to\gamma\gamma$ amplitude.  At $T=0$
the anomalous contribution to the divergence of the axial-vector vertex is
saturated by the pseudoscalar piece of the pion Bethe-Salpeter
amplitude\cite{mrpion}
\begin{eqnarray}
\nonumber
\lefteqn{\hat T_{\mu\nu}(k_1,k_2) = {\rm tr}\int_{l,q}^{\bar\Lambda}\,
S(q_1)\,\gamma_5 \tau^3 i E_\pi(\hat q;-P)\,}\\
& \times &
S(q_2)\,i{\cal Q}\Gamma_\mu(q_2,q_{12})\,
S(q_{12})\,i{\cal Q}\Gamma_\nu(q_{12},q_1)\,,
\label{pigganom}
\label{pi0gg}
\end{eqnarray}
where ${\cal Q}= {\rm diag}(2/3,-1/3)$ and herein $k_1=(\vec{k}_1,0)$,
$k_2=(\vec{k}_2,0)$, $P=k_1+k_2$, $q_1= q_{\omega_l}-k_1$, $q_2=
q_{\omega_l}+k_2$, $\hat q=\case{1}{2}(q_1+q_2)$, $q_{12}= q_{\omega_l} - k_1
+ k_2$.  Using Eq.~(\ref{bcvtx0}) to evaluate Eq.~(\ref{pi0gg}) for real
photons at $T=0$ one obtains
\begin{equation}
\label{tmunu}
\hat T_{\mu\nu}(k_1,k_2) = \frac{\alpha_{\rm em}}{\pi}\,
\epsilon_{\mu\nu\rho\sigma}\,k_{1\rho}\,k_{2\sigma}\,{\cal T}(0)\,,
\end{equation}
with in the chiral limit\cite{cdrpion}
\begin{equation}
f_\pi^0 \,{\cal T}(0) := g_{\pi^0\gamma\gamma}=1/2\,.
\end{equation}

At nonzero $T$ the tensor structure of Eq.~(\ref{tmunu}) survives to the
extent that, with our choice of $k_1$, $k_2$, it ensures one of the photons
is longitudinal (a plasmon) and the other transverse.  In this case we
determine the $T$-dependence using
\begin{equation} 
\hat T_{i4}(k_1,k_2) = \frac{\alpha_{\rm em}}{\pi}\,
(\vec{k_1}\times\vec{k_2})_i\,{\cal T}(0)
\label{anomT}
\end{equation}
and a generalisation of Eq.~(\ref{bcvtx0}) to nonzero $T$:
\begin{eqnarray}
\nonumber
\lefteqn{i \vec{\Gamma}(q_{\omega_{l_1}},q_{\omega_{l_2}})=
 \Sigma_A(q_{\omega_{l_1}}^2,q_{\omega_{l_2}}^2)\,i\vec{\gamma} }\\
&+& 
(\vec{q}_1+\vec{q}_2)\,
[\case{1}{2}\,i G(q_{\omega_{l_1}},q_{\omega_{l_2}})
        + \Delta_B(q_{\omega_{l_1}}^2,q_{\omega_{l_2}}^2)],\\
\nonumber \lefteqn{i\Gamma_4(q_{\omega_{l_1}},q_{\omega_{l_2}})=
\Sigma_C(q_{\omega_{l_1}}^2,q_{\omega_{l_2}}^2)\,i\gamma_4}\\
&& + (\omega_{l_1}+\omega_{l_2})\,
[\case{1}{2}\,i G(q_{\omega_{l_1}},q_{\omega_{l_2}})
        + \Delta_B(q_{\omega_{l_1}}^2,q_{\omega_{l_2}}^2)],\\
\nonumber \lefteqn{G(q_{\omega_{l_1}},q_{\omega_{l_2}}) =
\vec{\gamma}\cdot(\vec{q}_1 + \vec{q}_2)
\Delta_A(q_{\omega_{l_1}}^2,q_{\omega_{l_2}}^2) }\\
&& + \gamma_4 (\omega_{l_1} + \omega_{l_2})
        \Delta_C(q_{\omega_{l_1}}^2,q_{\omega_{l_2}}^2),
\end{eqnarray}
which satisfies the vector Ward-Takahashi identity
\begin{equation}
(q_{\omega_{l_1}}-q_{\omega_{l_2}})_\mu\,
i\Gamma_\mu(q_{\omega_{l_1}},q_{\omega_{l_2}})
= S^{-1}(q_{\omega_{l_1}}) - S^{-1}(q_{\omega_{l_2}}).
\end{equation}



\end{document}